\font\BBig=cmr10 scaled\magstep2


\def\title{
{\bf\BBig
\centerline{B\"acklund transformation}
\bigskip
\centerline{for}
\bigskip
\centerline{non-relativistic Chern-Simons vortices}
}
} 


\def\authors{
\centerline{
P.~A.~HORV\'ATHY and J.-C.~YERA}
\bigskip
\centerline{
D\'epartement de Math\'ematiques}
\medskip
\centerline{Universit\'e de Tours}
\medskip
\centerline{Parc de Grandmont,
F--37200 TOURS (France)
\foot{e-mail: horvathy@balzac.univ-tours.fr}
}
}

\def\runningauthors{
Horv\'athy \& Yera
}

\def\runningtitle{
B\"acklund transformation for Chern-Simons\dots
}


\voffset = 1cm 
\baselineskip = 14pt 

\headline ={
\ifnum\pageno=1\hfill
\else\ifodd\pageno\hfil\tenit\runningtitle\hfil\tenrm\folio
\else\tenrm\folio\hfil\tenit\runningauthors\hfil
\fi
\fi}

\nopagenumbers
\footline={\hfil} 


\def\and{\qquad\hbox{and}\qquad}

\def\kikezd{\parag\underbar} 
\def\IC{{\bf C}}

\def\smallover#1/#2{\hbox{$\textstyle{#1\over#2}$}}
\def\2{{\smallover 1/2}}
\def\ccr{\cr\noalign{\medskip}} 
\def\parag{\hfil\break} 
\def\={\!=\!}
\def\p{\partial}
\def\bp{\bar{\partial}}


\newcount\ch 
\newcount\eq 
\newcount\foo 
\newcount\ref 

\def\chapter#1{
\parag\eq = 1\advance\ch by 1{\bf\the\ch.\enskip#1}
}

\def\equation{
\leqno(\the\ch.\the\eq)\global\advance\eq by 1
}

\def\foot#1{
\footnote{($^{\the\foo}$)}{#1}\advance\foo by 1
} 

\def\reference{
\parag [\number\ref]\ \advance\ref by 1
}

\ch = 0 
\foo = 1 
\ref = 1 


\title
\vskip10mm
\authors
\vskip.25in

\parag
{\bf Abstract.}

{\it A B\"acklund transformation yielding the static
non-relativistic Chern-Simons vortices of 
Jackiw and Pi is presented}.

\vskip5mm
\noindent
(\the\day/\the\month/\the\year)

\medskip\noindent
Submitted to the {\sl Phy. Rev}. {\bf D}
\vskip5mm

\vskip5mm

\chapter{Introduction}

In the non-relativistic `Chern-Simons'
version of the Abelian Higgs model of
Jackiw and Pi [1-2-3], the scalar field,
$\Psi$, is described by the
{\it gauged, planar non-linear Schr\"odinger equation}
$$
i\partial_t\Psi=\big[
-{1\over2}(\vec{\nabla}-i\vec{A})^2+A^0
-g\,\Psi^*\Psi\big]\Psi.
\equation
$$
(We work in our units where $\hbar=m=e=1$). 
Here the `electromagnetic' field, associated with the 
vector potential $(A^0,\vec{A})$,
is assumed to 
satisfy the Chern-Simons field-current identity 
$$
B=\epsilon^{ij}\partial_iA^j=-{1\over\kappa}\varrho,
\qquad
E^i=-\partial_iA^0-\partial_tA^i={1\over\kappa}\,\epsilon^{ij}J^j,
\equation
$$
($i,j=1,2$), where
$
\varrho\=\Psi^*\Psi
$
and
$
\vec{J}\=(-i/2)\big[
\Psi^*\vec{D}\Psi-\Psi(\vec{D}\Psi)^*
\big]
$
denote the particle density and the current, respectively.
Explicit multivortex solutions
have only been found so far
for the special value of $g=\pm 1/\kappa$:
for static fields,
the second-order field equations above can be reduced to
the first-order `self-dual'  system
$
(D_x\pm iD_2)\Psi=0.
$ 
In a suitable gauge and away from the zeros of
$\varrho$, this becomes the Liouville equation
$$
\bigtriangleup\log\varrho=\pm{2\over\kappa}\varrho.
\equation
$$
\goodbreak\noindent
Regular solutions arise by chosing the upper (resp. lower) sign for 
$\kappa<0$ (resp. for $\kappa>0)$, 
$$
\varrho=4|\kappa|{|f'|^2\over\big[1+|f|^2\big]^2},
\equation
$$
where $f\equiv f(z)$ is 
a meromorphic function on the complex plane.
The remaining fields are expressed in terms of 
the particle density as
$$
A^0={1\over2|\kappa|}\varrho,
\qquad
\vec{A}=
{1\over2}\vec{\nabla}\times\log\varrho+\vec\nabla\omega,
\qquad
\Psi=\sqrt{\varrho}\,e^{i\omega},
\equation
$$
where $\omega$ is chosen so that $\vec{A}$ is regular at
the zeros of $\varrho$.

For $g\=\pm1/\kappa$, {\it all} 
static solutions are self-dual. This
can be proved using the conformal invariance of the system [2].

The problem of integrability of the full, time-dependent
second-order system (1.1-2) was examined by L\'evy et al. 
[4] and by Knecht et al. [5], 
who found that it was {\it  not} in general 
integrable. 
In [5] it was shown in particular, 
that the full system 
fails to pass the Painlev\'e test,
as extended to partial diffential equations by Weiss, Tabor and Carnevale
(WTC) [6].

The point is that the WTC method --- when it works ---
has the additional bonus to
provide B\"acklund transformations for generating solutions.
In this paper, we take advantage of this to construct,
in the static case and for 
$g\=\pm 1/\kappa$,
B\"acklund transformations allowing us
to rederive all static solutions.

\goodbreak
\chapter{A B\"acklund transformation}

Write $\Psi=\sqrt{\varrho}\,e^{i\omega}$
and introduce, following Knecht et al. [5], the new variables
$$\matrix{
\varrho=|\kappa|^3f^2,\qquad\hfill
&A^0-\partial^0\omega
=-\displaystyle{\kappa^2\over2}w\hfill
\ccr
A_1-\partial_1\omega
=-\kappa u,\hfill
&A_2-\partial_2\omega
=-\kappa v,\hfill
\ccr
x_1=\displaystyle{x\over|\kappa|}\hfill
&x_2=\displaystyle{y\over|\kappa|},
\cr}
\equation
$$
in terms of which the static CS Eqns read\foot{
One actually gets one more relation, which corresponds to the
continuity equation and appears here as a consistency condition.}
$$\eqalign{
u_y-v_x&=-f^2,
\cr
w_x&=2vf^2,
\qquad
w_y=-2uf^2,
\cr
f_{xx}+f_{yy}&=
-2\epsilon^{-1} f^3-f(w-u^2-v^2),
\cr}
\equation
$$
where $\epsilon=1/g|\kappa|$.
The WTC method [6], [5] amounts to developping 
the fields into a generalized Laurent series,
$$\matrix{
u=\displaystyle\sum_{k=0}^\infty u_k\Phi^{k-p_u},\hfill
&v=\displaystyle\sum_{k=0}^\infty v_k\Phi^{k-p_v},\hfill
\cr\cr
w=\displaystyle\sum_{k=0}^\infty w_k\Phi^{k-p_w},\hfill
&f=\displaystyle\sum_{k=0}^\infty f_k\Phi^{k-p_f},\hfill
\cr}
\equation
$$
where $\Phi=0$ is the `singular manifold'. Inserting these
expressions into the eqns. of motion
fixes the values of the $p$'s and 
provides us with recursion relations, except for some
particular values in $k$ called 'resonances', when consistency
conditions have to be satisfied. 
In detail, for $k=0$
we find, consistently with Knecht et al. [5],
that $p_u=p_v=p_f=1$, $p_w=2$
and
$$\matrix{
u_0=-\epsilon\,\Phi_y,\hfill
&v_0=\epsilon\,\Phi_x,\hfill
\ccr
w_0=\epsilon^2(\Phi_x^2+\Phi_y^2),\hfill
&f_0^2=-\epsilon\,(\Phi_x^2+\Phi_y^2).\hfill
\cr}
\equation
$$
Resonances occur for $k=1,2$ and $4$.
For $k=1$ we get 
$$
\eqalign{
2f_0f_1&=-{u_0}_y+{v_0}_x,
\cr
2f_0^2v_1+\Phi_xw_1+4v_0f_0f_1&={w_0}_x,
\cr
-2f_0^2u_1+\Phi_yw_1-4u_0f_0f_1&={w_0}_y,
\cr
-2f_0u_0u_1-2f_0v_0v_1+f_0w_1-6(\Phi_x^2+\Phi_y^2)f_1
&=
(\Phi_{xx}+\Phi_{yy})f_0+2\big(\Phi_x{f_0}_x+\Phi_y{f_0}_y\big),
\cr}
\equation
$$
>From the first of these equations we deduce, using Eq. (2.4)
$$
f_1^2=
-\epsilon\,{(\bigtriangleup\Phi)^2\over4(\Phi_x^2+\Phi_y^2)}.
\equation
$$
The remaining system of three equations has a vanishing determinant.
Consistency requires hence
$$
\epsilon^2=1
\qquad\hbox{i.e.}\qquad
g={1\over |\kappa|}.
\equation
$$
Then $u_1$ and $v_1$ can be expressed as a function of $w_1$,
which is arbitrary.

$k=2$ is again a resonance value: the l.h.s. of
$$\eqalign{
\Phi_yu_2-\Phi_xv_2+2f_0f_2
&=
-{u_1}_y+{v_1}_x-f_1^2,
\cr
2f_0^2v_2+4v_0f_0r_2
&=
{w_1}_x-2(f_1^2v_0+2v_1f_0f_1),
\cr
-2f_0^2u_2-4u_0f_0r_2
&=
{w_1}_y+2(f_1^2u_0+2u_1f_0f_1),
\cr
-2f_0u_0u_2-2f_0u_0v_2+f_0w_2
\quad&
\cr
\qquad
-6(\Phi_x^2+\Phi_y^2)f_2
&=
-({f_0}_{xx}+{f_0}_{yy})-6\epsilon^{-1}f_1^2f_0-w_1f_1
\cr
&\qquad
+f_0(u_1^2+v_1^2)+2f_1(u_0u_1+v_0v_1).
\cr}
\equation
$$
has vanishing determinant. Consistency requires hence the 
same to be true for the r.h.s., which only happens when 
$$
w_1=-2\epsilon\,f_0f_1
=-\bigtriangleup\!\Phi.
\equation
$$

The arisal of the constraint (2.9)
shows already that even the static system (2.2) 
fails to pass the Painlev\'e test of
WTC, which would allow an  arbitrary $w_1$. 
We can, nevertheless, continue our search for finding 
solutions constrained to satisfy (2.9).
Then $u_2$, $v_2$ and $w_2$ are expressed using the arbitrary
function $f_2$. 
(Condition (2.9) is below related to {\it self-duality}).

Inserting $w_1$ into (2.5)
we get, with the help of (2.4) and (2.6)
$$\eqalign{
&2\big(\Phi_x^2+\Phi_y^2\big)v_1
-\epsilon\,\bigtriangleup\Phi\,\Phi_x
=-\epsilon\big(\Phi_x^2+\Phi_y^2\big)_x,
\ccr
&2\big(\Phi_x^2+\Phi_y^2\big)u_1
+\epsilon\,\bigtriangleup\Phi\,\Phi_y
=\epsilon\big(\Phi_x^2+\Phi_y^2\big)_y,
\cr}
\equation
$$
while the last equation is identically satisfied.
Thus
$$\matrix{
u&=&-\epsilon\,(\log\Phi)_y&+&u_1&+&u_2\Phi+\dots,
\ccr
v&=&\epsilon\,(\log\Phi)_x&+&v_1&+&v_2\Phi+\dots,
\ccr
w&=&-\bigtriangleup\log\Phi&+&w_2&+&w_3\Phi+\dots,
\ccr
f^2&=&\epsilon\,\bigtriangleup\log\Phi&+&(2f_0f_2+f_1^2)&+
&(2f_1f_2+2f_0f_3)\Phi+\dots.
\cr}
\equation
$$

Now we try to {\it truncate} these infinite series by only keeping
terms of order less or equal to zero,
$$
u_k=v_k=w_{k+1}=f_k\equiv0
\quad\hbox{for}\quad
k\geq2.
\equation
$$
Inserting these relations into the first equation of (2.8),
we find that the  r. h. s. vanishes. Hence
$$
{u_1}_y-{v_1}_x=-f_1^2.
\equation
$$
>From the other eqns. of (2.8) we get, for the Ansatz (2.12) and 
using the constraint (2.9),
$$
{f_1}_x=-\epsilon v_1f_1,
\and
{f_1}_y=\epsilon u_1f_1
\equation
$$
and
$$
w_2=-\epsilon f_1^2.
\equation
$$
Note that if $f_1$ is not identically zero, then 
Eq. (2.14) implies
$$
\left({{f_1}_x\over f_1}\right)_x
+
\left({{f_1}_y\over f_1}\right)_y
=-\epsilon f_1^2
\equation
$$
which is the {\it Liouville equation}.
Requiring the constraint (2.9) means therefore
reducing the second-order equation (2.2) to a first-order system.

So far we have only considered  the terms $k=0,1,2$.
The consistency of our procedure follows from the verification,
using the formul{\ae} given in Ref. [5],
that all remaining equations, including the compatibily condition
for the resonance value $k=4$, are identically satisfied.
The case $k=3$ we shows in particular 
that the fields $f_1$, $u_1$, $v_1$, $w_2$ satisfy 
the  equations (2.2) we started with; they provide us therefore
with a ``seed solution'' in our B\"acklund transformation.

Collecting our results and returning to the physical variables,
cf. Eq. (2.1), we have proved the following. 
Let
$\rho\sim f_1^2,\, \vec{a}\sim (u_1,v_1)$,  $a^0\sim w_2$
be any ``seed solution'',
$$\kappa\,\vec\nabla\times\vec{a}
=-\rho,
\qquad
\vec\nabla\times\rho
=
-2\epsilon\,({\rm sign}\kappa)\,\vec{a}\,\rho,
\qquad
a^0=\epsilon{1\over2|\kappa|}\rho, 
\equation
$$
such that
$$\eqalign{
&(\bigtriangleup\Phi)^2
=-4{\epsilon\over|\kappa|}\,\rho\big[(\p_1\Phi)^2+(\p_2\Phi)^2\big],
\ccr
&2({\rm sign}\,\kappa)\big[(\p_1\Phi)^2+(\p_2\Phi)^2\big]
\,\epsilon_{ij}\,a_j
+\epsilon\bigtriangleup\!\Phi\,\p_i\Phi
=\epsilon\,\p_i\big[(\p_1\Phi)^2+(\p_2\Phi)^2\big]
\cr}
\equation
$$
($i=1,2$). Then the B\"acklund transformation
$$\left\{\eqalign{
&\varrho=\epsilon|\kappa|\,\bigtriangleup\log\Phi+\rho,
\ccr
&\vec{A}=
\epsilon\,({\rm sign}\,\kappa)\,\vec\nabla\times\log\Phi
+\vec{a}
+\vec\nabla\omega,
\ccr
&A^0={1\over2}\bigtriangleup\log\Phi+a^0
\cr}\right.
\equation
$$
(where $\varrho\sim f^2,\, \vec{A}\sim (u,v),\, A^0\sim w$)
provides us with a new set of solutions for (1.1-2).

This follows either from the proof above, or can 
be directly verified.
 
\goodbreak
If $\rho\neq0$, the first two equations in (2.17) reduce 
to the 
Liouville equation for $\rho$, while $\vec{a}$ and $a^0$ are
expressed as in (1.5). The seed solution is hence
necessarily {\it self-dual}.
\goodbreak

\goodbreak
\chapter{The construction of solutions}

The seed solution may {\it not} be a physical one:
a judicious choice may simplify a great deal the equations to be solved.
Below we obtain in fact the {\it general solution} (1.4)
by chosing $\rho$ to {\it vanish}. 
$$
\rho\equiv0
\equation
$$
In detail, let as assume that (3.1) holds. Then Eq. (2.17) 
is satisfied by $a^0\equiv0$ and $\vec{a}$ any curl-free vectorpotential.
Similarly, the upper equation in (2.18) now requires that $\Phi$ solve
the Laplace equation
$$
\bigtriangleup\Phi=0.
\equation
$$
Introducing the complex notations $z=x_1+ix_2$, $\bar{z}=x_1-ix_2$,
$\partial=\2(\partial_1-i\partial_2)$, 
$\bp=\2(\partial_1+i\partial_2)$, the solution of (3.2) is given by
$
\Phi(z,\bar{z})=f(z)+g(\bar{z}),
$
where $f(z)$ is analytic and $g(\bar{z})$ is anti-analytic.
With this choice, $\varrho$ is seen to be real
if $g(\bar z)=1/\overline{f(z)}$. Hence
$$
\Phi(z,\bar{z})=f(z)+{1\over\overline{f(z)}}.
\equation
$$
Positivity of $\varrho$ requires finally to set $\epsilon\=1$. 
In conclusion, the particle density is
$$
\varrho=
|\kappa|\,\bigtriangleup\log\Phi
=
|\kappa|\,
\bigtriangleup\log\big[1+|f|^2\big],
\equation
$$
i.e., the general solution  (1.4) of the Liouville eqn.

The second Eqn. in (2.18) allows us to express the vector potential of
the seed solution as
$$
\vec{a}=
-\2({\rm sign}\,\kappa)\,\vec\nabla\times
\log\big[(\p_1\Phi)^2+(\p_2\Phi)^2\big]
\equiv
-\2({\rm sign}\,\kappa)\,\vec\nabla\times\log\,
{|f'|^2\over\bar{f}^2},
\equation
$$
whose curl indeed vanishes, as required for consistency. 
Note that $\vec{a}$ combines with the first term in the 
$\vec{A}$-equation of (2.19)
to yield the curl of $\log\varrho$, cf. Eq. (1.5).
Finally, $A^0\=\varrho/2|\kappa|$ 
follows from (2.17) and (2.19).

For example, for each fixed $0\neq c\in\IC$,
$$
\Phi=(z/c)^N+(\bar{z}/\bar{c})^{-N}
\equation
$$
is a rotationally invariant solution of the Laplace
equation, which yields the well-known radial solution
$$
\varrho=
{4N^2|\kappa|\over r^2}\left(
\big({r\over r_0}\big)^N+\big({r_0\over r}\big)^N\right)^{-2}
\equation
$$
($r_0=|c|$ [1-3]).
The first term in the $\vec{A}$-equation of (2.19) is 
$$
({\rm sign}\,\kappa)\,i\,{2N\over\bar{z}}\,{1\over1+|z|^{2N}},
\equation
$$
while the seed solution is 
$$
a\equiv a_1+ia_2=
-({\rm sign}\,\kappa)\,i\,{N+1\over\bar{z}}.
\equation
$$
At the origin, the sum of these terms behaves as 
$$
({\rm sign}\,\kappa)\,i\,{(N-1)\over\bar{z}},
\equation
$$
so that the singularities are avoided 
if the phase $\omega$ ($\partial_t\omega=0$) is chosen to be
$$
\omega=(-{\rm sign}\,\kappa)\,(N-1)\,{\rm arg}\ z.
\equation
$$

The magnetic charge 
$Q\!\equiv\!\int\!B\, d^2\vec{r}$
is conveniently calculated as 
$$
Q=\oint_{S}\vec{A}\cdot d\vec\ell,
\equation
$$
where $S\equiv {S_\infty}$ denotes the circle at infinity.  
At infinity (3.8) falls off, so that 
only the seed and $\omega$ terms contribute.
We find
$$
Q=
-({\rm sign}\,\kappa)\,2\pi(N+1)-({\rm sign}\,\kappa)\,2\pi(N-1)
=-({\rm sign}\,\kappa)\,4\pi N,
$$ 
as expected.
More generally,
$$
\Phi={\prod_{i=1}^N\big(z-z_i)\over P(z)}
+
{\overline{P(z)}\over 
\prod_{i=1}^N\big(\bar{z}-\bar{z}_i)}
\equation
$$
where the $z_i$'s are arbitrary complex numbers 
and $P(z)$ is a polynomial of $z$ of degree at most $N-1$
($P(z_i)\neq0$), provides us with a $4N$-parameter family of
solutions with magnetic charge
$Q=(-{\rm sign}\,\kappa)4N\pi$.
This can be shown along the same lines as above.

\goodbreak
\chapter{Discussion}

Our formul{\ae} are equivalent to the
expression in Refs. [1-3], [7]. 
To see this, observe that $\Phi$ is
manifestly invariant with respect to the transformation
$f\to1/\bar{f}$. But $\varrho$ is also
invariant with respect to complex conjugation  
$f\to\bar{f}$ (cf. (3.4)), and thus also with respect to
$f\to1/f$. But for (3.13),
$1/f$ is decomposed into partial fractions as
$$
{1\over f}=\sum_{i=1}^N{d_i\over z-z_i}.
\equation
$$
which is the standard choice [1-3], [7].

A B\"acklund transformation for the Liouville equation  
has been constructed before by D'Hoker and Jackiw [8]. Their
approach also involves a solution of the Laplace equation.
They need, however, to solve an additional system of coupled, 
first-order differential equations.
Our formul{\ae} are hence different from theirs.

\goodbreak
\kikezd{Acknowledgements}. We are indebted to Professors 
R.~Jackiw and L.~V\'eron for discussions.
J.-C. Yera would like to thank the Government of La C\^ote d'Ivoire
for a doctoral scholarship and to the 
Mathematics Department of Tours University
for hospitality.

\vskip5mm
\goodbreak

\centerline{\bf\BBig References}

\reference
R.~Jackiw and S-Y.~Pi,
{\sl Phys. Rev. Lett}. {\bf 64}, 2969 (1990);
{\sl Phys. Rev}. {\bf D42}, 3500 (1990).

\reference
R. Jackiw and S-Y. Pi, 
{\sl Prog. Theor. Phys. Suppl}. {\bf 107}, 1 (1992).

\reference
G. Dunne, {\sl Self-Dual Chern-Simons Theories}. 
Springer Lecture Notes in Physics. New Series: Monograph 36. (1995).

\reference
D. L\'evy, L. Vinet, P. Winternitz,
{\sl Ann. Phys}. {\bf 230}, 101 (1994).

\reference
M. Knecht, R. Pasquier and J. Y. Pasquier,
{\sl Journ. Math. Phys}. {\bf 36}, 4181 (1995).
\goodbreak

\reference
J. Weiss, M. Tabor, and G. Carnevale,
{\sl J. Math. Phys}. {\bf 24}, 522 (1983).

\reference 
S. K. Kim, K. S. Soh, and J. H. Yee,
{\sl Phys. Rev}. {\bf D42}, 4139 (1990).

\reference
E.~D'Hoker and R.~Jackiw,
{\sl Phys. Rev}. {\bf D26}, 3517 (1982).

\bye